\documentclass[conference]{IEEEtran}
\IEEEoverridecommandlockouts
\usepackage{listings}
\usepackage{amsmath,amssymb,amsfonts}
\usepackage{graphicx}
\usepackage{tikz}
\usetikzlibrary{backgrounds,fit,decorations.pathreplacing,calc}
\usepackage{xcolor}
\usepackage{booktabs}
\usepackage[caption=false,font=footnotesize]{subfig}

\usepackage{blochsphere}
\def\BibTeX{{\rm B\kern-.05em{\sc i\kern-.025em b}\kern-.08em
    T\kern-.1667em\lower.7ex\hbox{E}\kern-.125emX}}
    
\makeatletter
\renewcommand*\env@matrix[1][\arraystretch]{%
  \edef\arraystretch{#1}%
  \hskip -\arraycolsep
  \let\@ifnextchar\new@ifnextchar
  \array{*\c@MaxMatrixCols c}}
\makeatother
    
\usepackage[pagebackref]{hyperref}
\usepackage{lipsum}

\usepackage{svg}

\definecolor{tdgreen}{rgb}{0,0.6,0}

\usepackage{tikz}
\usepackage{mathtools}
\usetikzlibrary{arrows.meta}

\usepackage{circuitikz}

\makeatletter
\pgfcircdeclarebipolescaled{instruments}
{
    \anchor{text}{\pgfextracty{\pgf@circ@res@up}{\northeast}
        \pgfpoint{-.5\wd\pgfnodeparttextbox}{
            \dimexpr.5\dp\pgfnodeparttextbox+.5\ht\pgfnodeparttextbox+\pgf@circ@res@up\relax
        }
    }
}
{\ctikzvalof{bipoles/oscope/height}}
{josephson}
{\ctikzvalof{bipoles/oscope/height}}
{\ctikzvalof{bipoles/oscope/width}}
{
    \pgf@circ@setlinewidth{bipoles}{\pgfstartlinewidth}
    \pgfextracty{\pgf@circ@res@up}{\northeast}
    \pgfextractx{\pgf@circ@res@right}{\northeast}
    \pgfextractx{\pgf@circ@res@left}{\southwest}
    \pgfextracty{\pgf@circ@res@down}{\southwest}
    \pgfmathsetlength{\pgf@circ@res@step}{0.25*\pgf@circ@res@up}
    \pgfscope
        \pgfpathrectanglecorners{\pgfpoint{\pgf@circ@res@left}{\pgf@circ@res@down}}{\pgfpoint{\pgf@circ@res@right}{\pgf@circ@res@up}}
        \pgf@circ@draworfill
    \endpgfscope
    \pgfscope
      \pgfpathmoveto{\pgfpoint{\pgf@circ@res@left}{\pgf@circ@res@up}}%
      \pgfpathlineto{\pgfpoint{\pgf@circ@res@right}{\pgf@circ@res@down}}%
      \pgfpathmoveto{\pgfpoint{\pgf@circ@res@right}{\pgf@circ@res@up}}%
      \pgfpathlineto{\pgfpoint{\pgf@circ@res@left}{\pgf@circ@res@down}}%
      \pgfusepath{draw}
    \endpgfscope
}
\def\pgf@circ@josephson@path#1{\pgf@circ@bipole@path{josephson}{#1}}
\tikzset{josephson/.style = {\circuitikzbasekey, /tikz/to path=\pgf@circ@josephson@path, l=#1}}

\usetikzlibrary{quantikz}

\usepackage[frozencache,cachedir=.]{minted}

\begin{document}

%to suppress long author lists in IEEE (from: https://tex.stackexchange.com/questions/164017/limiting-the-number-of-authors-in-the-references-with-ieeetran)
\bstctlcite{IEEEexample:BSTcontrol}

\title{Quantum Computing 2022\thanks{Corresponding author: \url{james.d.whitfield@dartmouth.edu}}}

% \author{
% \IEEEauthorblockN{James D. Whitfield,
%   Jun Yang, Weishi Wang, Joshuah T. Heath, and  Brent Harrison}
% \IEEEauthorblockA{\textit{Department of Physics and Astronomy, Dartmouth College} \\
% Hanover, New Hampshire, USA
% %Email: james.d.whitfield@dartmouth.edu
% }
% \IEEEauthorblockA{\textit{AWS Center for Quantum Computing} \\
% Pasadena, California, USA
% %Email: james.d.whitfield@dartmouth.edu
% }
% }
\author{\IEEEauthorblockN{James D. Whitfield\IEEEauthorrefmark{1}\,\IEEEauthorrefmark{2},
Jun Yang\IEEEauthorrefmark{1}, 
Weishi Wang\IEEEauthorrefmark{1}, 
Joshuah T. Heath\IEEEauthorrefmark{1}, 
and Brent Harrison\IEEEauthorrefmark{1}  }
\IEEEauthorblockA{\IEEEauthorrefmark{1}
Department of Physics and Astronomy,
Dartmouth College\\ Hanover, New Hampshire, USA 03755}
\IEEEauthorblockA{\IEEEauthorrefmark{2}AWS Center for Quantum Computing \\
Pasadena, California, USA 91125}}

\pagestyle{plain}

\maketitle

\thispagestyle{plain}

\begin{abstract}
Quantum technology is full of figurative and literal noise obscuring its promise. In this overview, we will attempt to provide a sober assessment of the promise of quantum technology with a focus on computing. We provide a tour of quantum computing and quantum technology that is aimed to be comprehensible to scientists and engineers without becoming a popular account.  The goal is not a comprehensive review nor a superficial introduction but rather to serve as a useful map to navigate the hype, the scientific literature, and upcoming press releases about quantum technology and quantum computing.  We have aimed to cite the most recent topical reviews, key results, and guide the reader away from fallacies and towards active discussions in the current quantum computing literature.  The goal of this article was to be pedantic and introductory without compromising on the science. 
\end{abstract}

The field of quantum technology, especially quantum computing technology, has emerged as an active area of academic and corporate research and development. It has also emerged as an area of heavy investment by companies, governments, and private investors worldwide. For example, worldwide investment in 2021 was estimated at \$24.4 billion with the United States appropriating \$1.2 billion over five years through the National Quantum Initiative Act~\cite{qureca_2021, raymer2019us}. So far this global investment has paid off in rampant technological progress, flagship experiments, and major intellectual developments.

%\paragraph{Opening paragraph}
Quantum technologies are the detectors, devices, and communication systems that rely on uniquely quantum resources, and quantum computing is the use of these quantum technologies to speed up or otherwise improve solutions to computational tasks and problems. The encapsulation of quantum mechanics has yielded a number of quantum technologies useful for storing and manipulating quantum information. Taking the position that information is physical \cite{Landauer2008Jan}, the objects carrying or storing that information may exhibit uniquely quantum behavior and then we refer to the information stored as \emph{quantum information}.  If we consider the smallest unit of information as a bit taking values of zero or one, then, similarly, the smallest unit of quantum information is the qubit (or quantum bit).

%\paragraph{Outline}
In Section \ref{sec:qtech} an overview of key modern quantum technologies is given. The theory driving these devices is given in Section \ref{sec:qtheory}. Our presentation of quantum mechanics is somewhat novel in that we begin with the quantum state as a quantum probability density matrix extended directly from ordinary probability density vectors. These key ideas prepare for Section \ref{sec:gates} where we introduce the gate model of quantum computation and some equivalent formulations. In Section \ref{sec:primacy}, we discuss the claims and counter-claims of quantum primacy\footnote{We have chosen the term \emph{quantum primacy} rather than \emph{quantum supremacy} originally defined in~\cite{Preskill2012Mar}. Both terms are used in the literature to refer to the same idea. However, there has been some controversy and debate about this naming convention in the literature and popular science press~\cite{wiesner_physicists_nodate, palacios-berraquero_instead_2019, preskill_why_2019}. \emph{Quantum advantage} is usually reserved for quantum computers merely edging out conventional computers on tasks of practical interest. This idea is slightly different from the original idea of quantum supremacy / primacy where the computational separation should be distinctively large.}.
Algorithms for quantum computers are discussed in Section \ref{sec:applications} where we have provided an overview of the quantum algorithms literature, including the quantum singular value transformation framework.  In the last section, Section \ref{sec:outlook}, we give an outlook on quantum technology including next steps for the reader including some sample code.

\section{Quantum technologies}\label{sec:qtech}

%\paragraph{opening paragraph }
Conventional
computers%
\footnote{Our preference is to use the term `conventional' computer rather than `classical' computer since the production and fabrication of devices at the nanometer scale requires domain knowledge implicitly based on quantum mechanics even if quantum coherence is not used in the operation of the resulting device.} rely on semi-conductor technologies where electrical currents are the primary information carriers. Quantum computing technology is far more diverse in that there is a wider range of physical systems used as information carriers. In this section, we focus on technologies for realizing quantum information carriers that have received the most attention and traction. This section gives a high-level overview but
for a more comprehensive view of quantum computing technologies see \cite{Ladd2010Mar,heinrich2021quantum}. 

%\paragraph{aims of section}
We will begin with the most widely adopted technology thus far, namely superconducting qubits. Then we turn toward quantum computing systems where individual atoms are used as information carriers.  Although not yet commercially relevant, nuclear magnetic resonance (NMR) quantum computing is included for historical and pedagogical reasons. From there we give mention to quantum light (photons) used to store and manipulate quantum information. Finally, this section will close with a brief look at other technologies that may be relevant in the coming years.

%\paragraph{Superconducting qubits}
Superconducting qubits are the most widely available device architecture for quantum computing and have received the most commercial attention thus far \cite{Rasmussen2021,Kjaergaard2020Mar}. The core technology is the Josephson Junction \cite{Josephson1962Jul} consisting of two superconducting metals separated by an insulator.  The superconducting state \cite{Tinkham2004Jun} requires low temperatures provided by dilution refrigerators.  There are numerous ways to combine inductors, capacitors, and Josephson junctions into varying circuit designs \cite{blais2021circuit} and functionalities \cite{Kjaergaard2020Mar}. The transmon design \cite{Koch2007} (a circuit consisting of a Josephson junction in parallel with a capacitor) is widely used as part of different qubit architectures pursued at companies such as Amazon, D-Wave Systems, Rigetti Computing, Google, and IBM.

Now, superconducting quantum computing devices with over 50 interacting qubits are becoming increasingly available \cite{arute2019quantum,zhu_quantum_2021, wu_strong_2021}. The discussion of the flagship quantum primacy experiments conducted with these devices is collected in Section \ref{sec:primacy}. Presently, we continue with atom-based quantum computing platforms.

%\subsection{Atomic comps}
Atomic physics gave the earliest experimental evidence of quantum theory; thus, the use of atoms as quantum information carriers is not so surprising. Much more surprising, however, is the level of controllability that has been demonstrated to date \cite{Wineland2013,Haroche2013}. First, we consider ion trap quantum computing then neutral atom quantum computing. 

%\paragraph{ion trap comps}
Ion trap quantum computing manipulates microscopic crystals composed of a handful of ionized atoms~\cite{bruzewicz2019trapped}. Each ion is addressed optically to isolate and control individual qubits of quantum information. Vibrations within the micro-crystal allow for controllable interactions between the  ions~\cite{bruzewicz2019trapped,Monroe2021,Ryan-Anderson2021}. Although commercially available ion trap quantum computers are smaller than superconducting devices, the control over the quantum state is typically greater, allowing for experimental demonstration of quantum simulations \cite{Monroe2021} and quantum error correction \cite{Ryan-Anderson2021}.

%\paragraph{neutral atoms}
Neutral atoms are trapped and manipulated using optics. Optical tweezers use laser light to trap and arrange the atoms into a 2D array \cite{Wu2021Feb}. The atom-atom interactions are mediated using the Rydberg excitations of electrons. When an electron is excited into a Rydberg state, the electron is far from the atomic nucleus but nonetheless remains bound to the atom. An apt analogy of the excited electron in a Rydberg state is that of a very distant but still gravitationally bound satellite of earth e.g. Halley's comet. The large extent of the electron in the Rydberg state can be used to mediate the multi-atom interactions. Early experimental setups only allowed model condensed matter systems to be tested \cite{Ebadi2021Jul} but recent progress has resulted in far more general quantum computing devices \cite{Bluvstein2021Dec,graham2022demonstration}.

%\subsection{NMR}
In Nuclear Magnetic Resonance (NMR), the quantum spin of nuclei in a strong applied magnetic field is accessed and manipulated with microwave pulses~\cite{xin2018nuclear}. Although not widely pursued at the industrial level, NMR quantum computing has important conceptual significance. Many of the key quantum computing concepts were first uncovered in this context and many of the first experimental implementations were done using NMR quantum computing~\cite{cory1998nuclear, xin2018nuclear}.

%\subsection{Optical systems}
While light sources are important for manipulating superconductors and controlling atomic systems, quantum properties of light can also be used for computational gain. Each unit of light (a photon) has two orthogonal directions for its polarization which can be manipulated quantum mechanically. However, the photons must be emitted in a highly correlated state before manipulation~\cite{knill2001scheme}. This process of obtaining sufficiently correlated units of light is probabilistic, but can be heralded (i.e. a secondary flag is able to indicate a successful generation attempt).  By multiplexing entanglement generation trials, companies such as PsiQ~\cite{bartolucci2021creation} and Xanadu~\cite{bourassa2021blueprint} as well as national collaborations  are attempting to create quantum computers with light as the primary information carrier. Photons are also called `flying' qubits since they can be used to couple distant quantum devices for computational or cryptographic purposes. Unlike some other quantum information paradigms, cryogenic temperatures are not needed.

%\subsection{NV} 
Lastly, it is worth mentioning defect-based quantum computing. Here, defects in a crystal, like diamond, can be used to store and manipulate quantum information. Nitrogen vacancy defects in diamond have seen the most development, but other defects like silicon in diamond are also being explored \cite{Chatterjee2021Mar}. The implantation of these vacancies as well as isotopic purity are difficult to control and have prevented defect-based quantum computing devices from reliably storing large amounts of quantum information. To date, the primary technological significance of nitrogen vacancy centers has been in the area of quantum sensing applications \cite{zhang2020diamond}.

%\paragraph{transition}
Following this rapid survey of devices used in quantum technology, we now turn to the theory. We will strive to give the principles behind the quantum theory of these devices and of quantum computing more broadly. We will begin with the probability-first approach to quantum mechanics \cite{Whitfield20kinematics}. Our introduction to quantum technology is aimed to allow engaged and knowledgeable readers opportunities to incorporate their own expertise. Similarly, our approach to quantum theory will allow readers to extend their analytical understanding of probability theory to quantum theory. This is a novel approach as compared to standard introductions e.g. \cite{nielsen2002quantum}. 

%A more direct and mathematical presentation can be found in q101

\section{Quantum theory}\label{sec:qtheory}

\begin{figure}
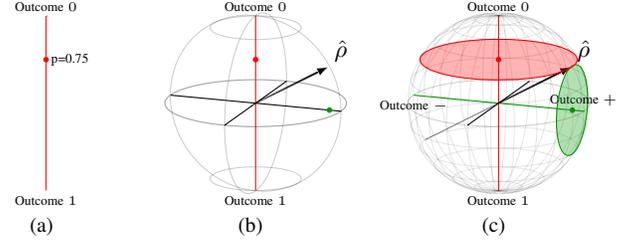

    \centering
\subfloat[]{
% prob line
\begin{blochsphere}[radius=1.2cm,tilt=15,rotation=-20,ball=none,opacity=.9]

% \drawBallGrid[style={opacity=.1}]{15}{15}

%z axis
\labelLatLon{up}{90}{0}
\labelLatLon{down}{-90}{90}

%x axis
\labelLatLon{left}{0}{0}
\labelLatLon{right}{0}{180}

%special angles
\labelLatLon[scale=.5]{thirtyup}{90}{0}
\labelLatLon[scale=.7]{fortyfiveup}{90}{0}
\labelLatLon[scale=.87]{sixtyright}{0}{0}
 
%label on the line
\node[circle,fill,red,scale=.25] at (thirtyup) {*};
\node[right] at (thirtyup) {{\tiny p=0.75}};
% \node[circle,fill,tdgreen,scale=.25] at (sixtyright) {*};

%eigenaxis for state rho
% \drawAxis[style=gray,opacity=0]{60}{0}

\node[above] at (up) {{\tiny Outcome $ 0$}};
\node[below] at (down) {{\tiny Outcome $ 1$}};

% \node[right] at (left) {{\tiny Outcome $ + $}};
% \node[left] at (right) {{\tiny Outcome $ - $}};

%\labelLatLon{psix}[scale=0.5]{0}{90}
%\labelLatLon{psiz}[scale=0.5]{90}{0}

%\drawLatitudeCircle[rotation=-160,tilt=45,style={tdgreen, fill, fill opacity=.25}]{45}

% x,z - axis
\drawAxis[style=red]{0}{0}
%\drawAxis[style=tdgreen,opacity=0]{90}{0}

%dot
%\drawBall[ball=3d,color=blue]

\end{blochsphere}
}
\subfloat[]{
\begin{blochsphere}[radius=1.2cm,ball=none,opacity=1,axisarrow=]
\drawBallGrid[style={opacity=.2}]{60}{90}
%\drawBall[ball=3d,opacity=.1]
%z axis
\labelLatLon{up}{90}{0}
\labelLatLon{down}{-90}{90}
%
%x axis
\labelLatLon{left}{0}{0}
\labelLatLon{right}{0}{180}
%
%special angles
\labelLatLon[scale=.5]{thirtyup}{90}{0}
\labelLatLon[scale=.7]{fortyfiveup}{90}{0}
\labelLatLon[scale=.87]{sixtyright}{0}{0}
%
% %states
% % Psi
% \drawStateLatLon{psi}{45}{0}
% \node[above right] at (psi) {{\tiny $\vec\psi $}};
%
% % x,y projections psi
% \drawSmallCircle[style={tdgreen,fill,fill opacity=.1}]{90}{0}{45}
% \drawLatitudeCircle[style={red, fill, fill opacity=.1}]{45}
%
% %label on the line
% \node[circle,fill,red,scale=.25] at (fortyfiveup) {*};
%
% %eigenaxis for state psi
%\drawAxis[style=gray,opacity=0]{60}{180}
%
%Rho
\drawStateLatLon{rho}{30}{0}
\node[above right] at (rho) {{\small $\hat\rho$}};
%
% x,y projection rho
%\drawSmallCircle[style={tdgreen,fill,fill opacity=.3}]{90}{0}{60}
%\drawLatitudeCircle[style={red, fill, fill opacity=.3}]{30}

%label on the line
\node[circle,fill,red,scale=.25] at (thirtyup) {*};
%\node[right] at (thirtyup) {{\tiny p=0.75}};
\node[circle,fill,tdgreen,scale=.25] at (sixtyright) {*};

%eigenaxis for state rho
% \drawAxis[style=gray,opacity=0]{60}{0}

\node[above] at (up) {{\tiny Outcome $ 0$}};
\node[below] at (down) {{\tiny Outcome $ 1$}};

% \node[right] at (left) {{\tiny Outcome $ + $}};
% \node[left] at (right) {{\tiny Outcome $ - $}};

%\labelLatLon{psix}[scale=0.5]{0}{90}
%\labelLatLon{psiz}[scale=0.5]{90}{0}

%\drawLatitudeCircle[rotation=-160,tilt=45,style={tdgreen, fill, fill opacity=.25}]{45}

% x,z - axis
\drawAxis[style={red}]{0}{0}
\drawAxis[style={black}]{90}{90}
\drawAxis[style={black}]{90}{0}
%\drawAxis[style=tdgreen,opacity=0]{90}{0}

%dot
%\drawBall[ball=3d,color=blue]

\end{blochsphere}
}
\subfloat[]{
\begin{blochsphere}[radius=1.2cm,tilt=15,rotation=-20,ball=none,opacity=.9]

\drawBallGrid[style={opacity=.1}]{15}{15}

%z axis
\labelLatLon{up}{90}{0}
\labelLatLon{down}{-90}{90}

%x axis
\labelLatLon{left}{0}{0}
\labelLatLon{right}{0}{180}

%special angles
\labelLatLon[scale=.5]{thirtyup}{90}{0}
\labelLatLon[scale=.7]{fortyfiveup}{90}{0}
\labelLatLon[scale=.87]{sixtyright}{0}{0}

% %states
% % Psi
% \drawStateLatLon{psi}{45}{0}
% \node[above right] at (psi) {{\tiny $\vec\psi $}};

% % x,y projections psi
% \drawSmallCircle[style={tdgreen,fill,fill opacity=.1}]{90}{0}{45}
% \drawLatitudeCircle[style={red, fill, fill opacity=.1}]{45}

% %label on the line
% \node[circle,fill,red,scale=.25] at (fortyfiveup) {*};

% %eigenaxis for state psi
%\drawAxis[style=gray,opacity=0]{60}{180}

% Rho
\drawStateLatLon{rho}{30}{0}
\node[above right] at (rho) {{\small $\hat\rho$}};
%\node at (rho) {{\tiny $\hat\rho$}};

% x,y projection rho
\drawSmallCircle[style={tdgreen,fill,fill opacity=.3}]{90}{0}{60}
\drawLatitudeCircle[style={red, fill, fill opacity=.3}]{30}

%label on the line
\node[circle,fill,red,scale=.25] at (thirtyup) {*};
\node[circle,fill,tdgreen,scale=.25] at (sixtyright) {*};

%eigenaxis for state rho
\drawAxis[style=gray,opacity=0]{60}{0}

\node[above] at (up) {{\tiny Outcome $ 0$}};
\node[below] at (down) {{\tiny Outcome $ 1$}};

\node[above] at (left) {{\tiny Outcome $ + $}};
\node[below] at (right) {{\tiny Outcome $ - $}};

%\labelLatLon{psix}[scale=0.5]{0}{90}
%\labelLatLon{psiz}[scale=0.5]{90}{0}

%\drawLatitudeCircle[rotation=-160,tilt=45,style={tdgreen, fill, fill opacity=.25}]{45}

% x,z - axis
\drawAxis[style=red]{0}{0}
\drawAxis[style=tdgreen,opacity=0]{90}{0}

\drawAxis[style=black]{90}{90}

%dot
%\drawBall[ball=3d,color=blue]

\end{blochsphere}
}
    \caption{In sub-figure (a), we visualize the $p$, the probability of obtaining Outcome $0$, as a point along a line from zero to one. This picture is generalized as a Bloch sphere in sub-figure (b) used to visualize quantum probability distributions. Quantum probability distributions, $\hat\rho$, are vectors within the unit sphere. In sub-figure (c), the projection of the state $\hat\rho$ in two different measurement bases with the red axis corresponding to measurement in the $Z$ basis. The two outcomes along the green axis are canonically labelled as Outcomes $\pm$ and the project in green gives the probability of obtaining Outcome $+$ in the $X$ direction.
    }
    \label{fig:qstates}
\end{figure}

%\paragraph{probability states}
For our introduction to quantum theory, we will begin with the notion of a probability density vector $\vec p$ with components $p_i$ that correspond to the probability of the $i$th event occurring upon measurement, see Fig.~\ref{fig:qstates}. Note that any concept of measurement compatible with probability will also be compatible with the notion of quantum states discussed below.  Probability vectors must satisfy the following three properties to maintain a sane interpretation of probability~\cite{grinstead2012introduction}:
\begin{IEEEeqnarray}{lr}
\IEEEyesnumber\label{eq:probs}
    \textrm{Normalization:} &\|\vec p\|_1 =\textstyle{\sum_i| p_i|} =1\IEEEyesnumber\IEEEyessubnumber\\
    \textrm{Real valued:} &p_i \textrm{ is a real number }\IEEEyessubnumber\\
    \textrm{Positive semi-definite:   }& p_i \geq 0\IEEEyessubnumber
\end{IEEEeqnarray}
To derive possible stochastic transformations, these mathematical properties constrain what is possible. Pursing probability theory by considering these constraints without concern for their origins is the essence of a kinematic approach. We will take this approach to quantum after introducing the quantum constraints extending Eq.~\eqref{eq:probs}.

%\paragraph{quantum states}
For quantum states, we can analogously define the quantum probability density matrix, $\hat \rho$, as the generalization of the probability density vector with the following properties:
\begin{IEEEeqnarray}{lr}
\IEEEyesnumber\label{eq:qstate}
        \textrm{Normalization:} & \|\hat\rho\|_1 
                                 =\mathrm{Tr} |\hat\rho| ={\textstyle{ \sum_i } |\hat\rho_{ii}| =
                                 1}\IEEEyesnumber\IEEEyessubnumber\\
    \textrm{Real-valued spectrum:} & \hat\rho=\hat\rho^\dag\IEEEyessubnumber\\
        \textrm{Positive semi-definite:   } & \vec x^\dag \; \hat\rho \; \vec x\geq0 \textrm{ for all } \vec x\IEEEyessubnumber
\end{IEEEeqnarray}
Here, $\dagger$ indicates the conjugate transpose whereby $(A^\dag)_{pq}=A_{qp}^*$ and $(a+bi)^*=a-bi$ with $i^2=-1$. These mathematical extensions ensure that the diagonal of the quantum state, no matter how the matrix is represented or measured, remains a valid probability density vector. The kinematics of the quantum states then follow directly from maintaining these constraints to preserve a proper interpretation of the quantum state \cite{Whitfield20kinematics}.

%\paragraph{quantum measurement from prob theory}
The notion of measurement is inherited from one's taste in probability. For instance, when considering quantum theory applied to estimating outcomes of an experiment, a Bayesian approach may be most appropriate. But in the case of repeatedly measuring the state of a reproducible state, a frequentist interpretation may be more apt. In principle, the interpretation of measurements presents no more of a foundational concern to quantum theory than it does already in probability theory. The key difference is that the probability density matrix may appear differently depending on the set of states being used to represent (and to measure) the state. 

%\paragraph{qubits versus bits }
For the smallest comparative example, consider two outcomes: Outcome 0 and Outcome 1. The probability of getting Outcome 0 is given by a real number between values zero and unity. This is represented by the red line in Fig.~\ref{fig:qstates}. The possible quantum states (according to the constraints above) are all possible vectors within the sphere. The projection of the quantum state onto the probability axis gives the probability of obtaining an outcome corresponding to the event labelled at the positive endpoint of the probability line.  

%\paragraph{rotation of the basis, differing measurement outcomes}
In quantum theory, we can select the orientation of the measurement as any line of probability through the Bloch sphere. For probability theory, the only allowed change of the measurement orientation is a permutation of the events. This would correspond to considering the probability of realizing Outcome 1 instead of Outcome 0. Those fluent with matrix analysis \cite{horn_matrix_2012}, will recognize that permutations are a discrete subset of the continuous orthogonal group which is itself a subset of the unitary group of transformations.

%example unpacking
As an example of changing the quantum measurement orientation, consider the green probability line along the $X$-axis in Fig.~\ref{fig:qstates}c has two Outcome $+$ or Outcome $-$. This new line of probability corresponds to obtaining the Outcome $+$. For any qubit state, we can obtain the outcome probability using the projection of the state on the green axis. An example is depicted by the dot placed along the green axis.  

%passive v active rotations
It is useful to note that the rotation of the state (with a fixed measurement orientation) and the rotation of the measurement basis (with a fixed state) are effectively the same.  In the language of vector kinematics, the two types of rotations are termed active and passive rotations respectively \cite{Thornton04}. In quantum physics, these two types of rotations are termed the Schr\"odinger and Heisenberg pictures \cite{Griffiths2018Aug}.

%\paragraph{example of change of basis}
For a concrete example of changing the measurement basis, consider the Hadamard transformation as given in Table~\ref{tbl:HPTCNOTgates}. This operation rotates the $Z$-axis of the sphere by $90^\circ$ to obtain states in the $X$-$Y$ plane. The probability of obtaining outcome zero or one is half since all states in the $X$-$Y$ plane project to the midpoint of the $Z$ probability axis.  The Hadamard transform plays a special role in many quantum algorithms for its use in preparing quantum states with high coherence. 

%\paragraph{linear algebra, eigenbasis}
From linear algebra \cite{Horn2012Oct}, we know that there always exists a basis for measurement in which the quantum probability density matrix is diagonal and, consequently, is just an ordinary probability density vector. This basis is called the eigenbasis and such a basis always exists for probability density matrices. The eigenbasis is the measurement basis in which coherences are not necessary to describe the state. The basis states for this measurement are called eigenstates and the probabilities are called eigenvalues. In this measurement basis, there is no essential difference between a probability density vector and a probability density matrix other than the organization of the probability values into a vector or a diagonal matrix.

%\paragraph{coherence  and decoherence}
The off-diagonal elements of the quantum probability density matrices are called coherences and are allowed to take values from the complex field of numbers so long as the state remains consistent with the kinematic constraints of Eq.~\eqref{eq:qstate}.  Quantum theory differs from ordinary probability theory through the use of these off-diagonal degrees of freedom (called coherences) that are generated after certain quantum operations. These coherences disappear as the system's behavior becomes less quantum and more akin to standard probability theory. The process of losing coherence is defined as  {decoherence} and is the major stumbling block for technological realizations of quantum devices. Decoherence in experimental quantum devices often appears due to device control errors and unwanted interactions with the environment.  

%\paragraph{Entanglement} 
The discussion of quantum theory has thus far focused only on the kinematic differences between quantum and probability theory. However, the critical difference between quantum and probability theory is how correlations are treated. In ordinary probability, if we consider a probability density vector associated with two random variables, say $\vec p_{AB}$, then we can say the variables are correlated if $\vec p_{AB}\neq \vec p_A \times \vec p_B$. That is, if we cannot write the joint density as the product of two different densities. In quantum mechanics, the same idea of correlated states leads to definitions of the type: $\hat \rho_{AB} \neq \hat \rho _A \otimes \hat \rho_B$ where we have used the appropriate generalization of the Cartesian product for quantum probability density matrices known as the Kronecker tensor product.

Entanglement is a key distinguishing feature of quantum computation.  Ordinary probability allows for correlated variables, but the use of coherence allows for the two systems $A$ and $B$ to be more strongly correlated than is otherwise possible with two probabilistic systems. This excess correlation available to quantum systems is called entanglement \cite{Brandao2016Apr,Cirac2021rmp, Verstraete2008Mar}. An important example of an entangled state is the Bell state defined and discussed further in Fig.~\ref{fig:bell}.

\begin{figure}
    \centering
    \[
   \hat\rho_{Bell}=\hat\rho_{AB} =\frac12 \begin{pmatrix}
    1 & 0 & 0 & 1\\
    0 & 0 & 0 & 0\\
    0 & 0 & 0 & 0\\
    1 & 0 & 0 & 1
    \end{pmatrix}\neq \hat\rho_A \otimes \hat\rho_B
    \]
    \begin{tabular}{cc}
    If Alice obtains... & ... Bob will never obtain\\
    \hline
    0 & 1\\
    1 & 0\\
    + & $-$\\
    $-$ & +
    \end{tabular}
    \caption{There are two bases that Alice and Bob may measure: the 0-1 basis (on the $Z$-axis of probability shown in red) and the $\pm$ basis (measured along the $X$-axis of the Bloch sphere shown in green). After Alice has chosen her basis and obtained the outcome listed on the left column, Bob cannot obtain the outcome listed on the right no matter which basis he chooses to measure in. 
    }
    \label{fig:bell}
\end{figure}

%\paragraph{Consequences of entanglement}
Without high entanglement, quantum computers are not more powerful than conventional computation \cite{Vidal2003Oct} but it should be noted that entanglement is necessary but not sufficient for quantum computation (as per the matchgate and Clifford gate sets discussed in Section \ref{sec:gates}). The insights from the quantum information community have led to great advances in conventional computing through the study of matrix product states and more general tensor network states~\cite{Verstraete2008Mar,Cirac2021rmp,pan_solving_2021}. The key insight is that the compression and efficient manipulation of the quantum state becomes possible on conventional computers when the entanglement remains low~\cite{Verstraete2008Mar, Vidal2003Oct, Cirac2021rmp}. By contrast, highly entangled states can be used to demonstrate violations to the local realism hypothesis \cite{Einstein1935May} using Bell's inequality \cite{Bell1964Nov,Giustina2015}. Because of the violation of local realism, these highly-entangled states are also useful for communication security~\cite{broadbent2016quantum}.

\subsection{Quantum wave functions}
%\paragraph{What about wave functions and the Schrodinger equation}
In this brief introduction to quantum theory, we have not yet mentioned the Schr\"odinger equation nor wave functions. Instead, we have focused on quantum probability density matrices and their kinematics. If we followed historical and standard approaches, we would have begun with wave functions, linear algebra, and differential equations. The notion of quantization might have arisen as a consequence of solving a linear second-order differential equation.  Note that superposition and quantization feature in sound and heat waves, in classical electromagnetic theory, and many connected mathematical situations that are not related to quantum mechanics. In order to keep the discussion focused on the essentials of quantum theory, we instead mention wave functions as a special case of the quantum probability density matrices, so called pure states. 

%\paragraph{technical description of wave functions}
Wave functions\footnote{Here we use standard terminology although we only consider wave functions as vectors, i.e. $\vec \psi$, rather than functions, i.e. $\psi(x)$.} are quantum probability densities corresponding to a quantum state with deterministic outcome in at least one measurement direction. That is, there is a measurement direction where the probabilistic outcomes are describable as $\vec p_1 = (1,0,0,0,...0)^T$ corresponding to vectors of unit length in Fig.~\ref{fig:qstates}. In this case, the quantum probability density matrix is called pure and satisfies the projector condition: $\hat\rho^2=\hat\rho$.   In the case of a pure quantum state, the quantum density matrix simplifies and can be characterized by a single vector $\vec\psi$, called a wave function\footnote{Note that when there is a basis where $\vec p_1$ appears, then the quantum probability density matrix has only one eigenvector corresponding to the lone non-zero eigenvalue. This allows us to write $\hat\rho=\vec\psi \vec \psi^\dag$ with $\vec\psi$ defining the wave function.}. Note that in every measurement basis except the eigenbasis of the probability density matrix there will be coherences. As these coherences are lost with respect to any other basis, the state will no longer be describable by a single vector $\vec\psi$ and the correct description is given by the quantum probability density matrix formalism.  

%\paragraph{when the wave function formalism is useful}
In the next section, the choice between the wave function vectors and the more general quantum probability density matrices is not forced. This is because most quantum computing gate operations are considered to be noiseless unitary transformations. In the case of a single qubit, the unitary transformations correspond to rotations of the Bloch sphere vector without changing its length. More technically, the unitary matrices do not change the eigenvalues of the quantum probability density matrix. Consequently, if the quantum state is pure and there is only one non-zero eigenvalue, unitary operations will not decrease (nor increase) the purity. 

\subsection{Quantum states spaces compared to probability event spaces}
In deterministic computing, the probabilities are enforced to be either zero or one; however, to extend this notion to quantum information theory, it is useful to think of the bit as taking values with some probability $p$. By beginning with this outlook, we can avoid misleading arguments about the size of the quantum state space.
Regardless of whether we have $n$ bits or $n$ qubits, we have $2^n$ possible basis states. In the case of $n$ bits, we can describe the full state with a probability vector containing $2^n$ real-valued entries. For example, when $n=3$
\begin{equation}
\vec p =\begin{pmatrix}
prob(X=000)\\
prob(X=001)\\
prob(X=010)\\
prob(X=011)\\
prob(X=100)\\
prob(X=101)\\
prob(X=110)\\
prob(X=111)
\end{pmatrix}
\end{equation}
Due to the normalization condition, at most $2^n-1$ real values are needed.
The quantum system, by contrast, cannot be described by a probability vector alone. Nonetheless, upon measurement, for $n$ qubits there will be only $2^n$ possible outcomes. Thus, the distinction between $n$ qubits and $n$ bits is deeper than a simple counting argument.  

%\paragraph{closing/outlook}
The key takeaway from this section on theory is that quantum coherence and quantum entanglement are the essential resources that quantum devices exploit for computational gains. One frequently encounters misleading claims about superposition or about the size of the quantum state space as reasons for quantum computational gain. By beginning with and comparing against probability theory rather than classical mechanics, we hope it is clear that superposition and counting arguments do not account for the power of quantum computation. Moreover, this probability-first approach has allowed us to sidestep philosophical discussion of measurement and proceed to the essential difference between probability and quantum theory. Armed with these basic ideas of quantum theory, we now turn toward computation beginning with the models of quantum computing in the next section.

\section{Models of quantum computing}\label{sec:gates}
%\paragraph{Opening}
In this section, we will consider the models of computation used to design quantum algorithms. There are many excellent introductions to the quantum circuit model e.g.~\cite{nielsen2002quantum,qiskit_textbook}, so
we will be satisfied with a brief survey of the main ideas and key notions. This prepares us for a look at restricted quantum circuit models which are efficiently simulated without quantum resources. This connects to the central research questions of what is the difference between quantum and conventional computing.  We then close this section with a mention of some alternative models of quantum computation that are computationally equivalent to the quantum circuit model.  

%gates + Bloch sphere
Before delving into the circuit model, we will make two remarks to connect to the prior section. First, for one qubit, all quantum gates from the quantum circuit model can be visualized as transformations of the Bloch sphere. If the gates are performed perfectly then they correspond to rotations of the Bloch sphere. In more general transformations, the Bloch sphere may be deformed such that the length of the vector representing $\hat \rho$ can be less than unit length. (Recall, from Fig.~\ref{fig:qstates}, that a vector of length zero would correspond to the mid-point of the probability axis i.e. $p=0.5$.)

%wave functions
Second, we can consider moving away from the general picture of a gate or series of gates, $U$, acting on a quantum probability density matrix: $\mathcal{E}_U(\hat\rho)=\hat\rho_{output}$. If the circuit $U$ is done without any errors, then we can write,  $\mathcal{E}_U(\hat\rho)=U \hat\rho  U^\dag$. Further, if the probability density matrix is pure $\hat\rho^2 = \hat\rho$, then we may write $\hat\rho=\vec\psi \vec\psi^\dag$. Combining these two statements, we have $\mathcal{E}_U(\psi\vec\psi^\dag)=U\vec\psi (U\vec\psi)^\dag$ allowing the quantum dynamics to be describable as $U\vec\psi=\vec\psi_{output}$. Thus, in the case that both the gates are perfectly implemented and the probability density matrix is pure, we can use the wave function picture of $U\psi$ found in standard introductions \cite{nielsen2002quantum,qiskit_textbook}. This distinction will not be relevant for the remainder of the review.

%\paragraph{universal gate sets and reversibility}
For conventional computing in the circuit models, the NAND gate is universal for computation. This means that every other circuit has an equivalent construction using exclusively NAND gates. The NAND gate takes two inputs and returns a single output 0 if and only if both inputs are 1. Because two bits are used for input and only one bit for output, this gate is irreversible and cannot be used as a quantum circuit element in a straightforward way. The Toffoli gate, as shown in Table \ref{tbl:HPTCNOTgates}, is a reversible gate that is universal for conventional computation. Its existence as a quantum circuit element shows that all conventional computations can be converted to quantum computation. A few examples of quantum circuit elements are given in Table~\ref{tbl:HPTCNOTgates}. 

\begin{table}[t!]
\centering
\caption{Some standard quantum gates and their representations.}
    \label{tbl:HPTCNOTgates}

\begin{tabular}{lcc} %{p{2cm}p{2cm}p{2.5cm}}
\toprule
Operator & Gate & Matrix\\
\midrule
Hadamard ($\mathbf{H}$) &
{
$\begin{tikzcd}[thin lines]
\qw & \gate{H} &\qw   
\end{tikzcd} $
}
& 
$\frac{1}{\sqrt{2}}\begin{pmatrix}
    1 & 1 \\ 1 &-1
    \end{pmatrix}$ 
\\
Phase $(\mathbf{S})$  &
$\begin{tikzcd}[thin lines]
\qw &  \gate{S} & \qw 
\end{tikzcd}$
& $\begin{pmatrix} 
    1 & 0 \\ 0 & i
    \end{pmatrix}$ \\
\begin{tabular}{l}
     Controlled Not\\
$\mathbf{CNOT}$
\end{tabular}
&
$
\begin{tikzcd}[thin lines]
\lstick{} &  \ctrl{1} & \qw \\
\lstick{}  & \targ{} & \qw
\end{tikzcd}$
& $\begin{pmatrix} 1 & 0 & 0 & 0 \\ 0 & 1 & 0 & 0 \\ 0 & 0 & 0 & 1 \\ 0 & 0 & 1 & 0\end{pmatrix}$\\
$\pi / 8$ ($\mathbf{T}$) & 
$\begin{tikzcd}[thin lines]
\qw &  \gate{T} & \qw  
\end{tikzcd}
$
& $\begin{pmatrix} 
    1 & 0 \\ 0 & e^{i\pi / 4}
    \end{pmatrix}$ \\
\midrule
Toffoli gate  &
$
\begin{tikzcd}[thin lines,row sep=0.2cm]
\lstick{} &  \ctrl{1} & \qw \\
\lstick{} &  \ctrl{1} & \qw \\
\lstick{}  & \targ{} & \qw
\end{tikzcd} 
$
& 
% { 
\begingroup
\setlength\arraycolsep{2pt}
$\begin{pmatrix} %[.5] 
1 & 0 & 0 & 0 & 0 & 0 & 0 & 0\\
0 & 1 & 0 & 0 & 0 & 0 & 0 & 0\\ 
0 & 0 & 1 & 0 & 0 & 0 & 0 & 0\\ 
0 & 0 & 0 & 1 & 0 & 0 & 0 & 0\\
0 & 0 & 0 & 0 & 1 & 0 & 0 & 0\\
0 & 0 & 0 & 0 & 0 & 1 & 0 & 0\\ 
0 & 0 & 0 & 0 & 0 & 0 & 0 & 1 \\ 
0 & 0 & 0 & 0 & 0 & 0 & 1 & 0 
\end{pmatrix}$
\endgroup
% }
\\
\midrule
Pauli $X$ & %$\begin{tikzcd}[thin lines]\qw &  \gate{X} & \qw \end{tikzcd}$
&$\begin{pmatrix}0&1\\1&0\end{pmatrix}$\\
Pauli $Y$ &% \changeme{Y}
&$\begin{pmatrix}0 &-i\\i&0\end{pmatrix}$\\
Pauli $Z$ & %\changeme{Z} 
&$\begin{pmatrix}1 &0\\0&-1\end{pmatrix}$\\
Rotation about $\vec {\mathbf{n}}$ & 
$\begin{tikzcd}[thin lines]
\qw &  \gate{R_\mathbf{n}(\theta)} & \qw 
\end{tikzcd}$
& $\exp(i\theta\vec{\mathbf{n}}.(X,Y,Z)/2)$\\
\bottomrule
\end{tabular}

\end{table}

%\paragraph{universal quantum gate sets}
A universal gate set is a fixed set of operations that when composed together allow one to construct any arbitrary multi-qubit transformation. We note immediately that the number of gates need to construct an arbitrary unitary starting from a fixed set could potentially be very large. Nonetheless, given single-qubit rotations, almost any two-qubit gates will suffice \cite{Lloyd1995universallogic}. There are a few special instances where the two-qubit gate is insufficient: for example, if the two-qubit gate is actually composed of two independent single-qubit gates. However, if an operation or matrix is chosen at random then almost surely it is sufficient for universal quantum computation \cite{deutsch1995universality,Lloyd1995universallogic}. 
Thus, a universal gate set is achievable through almost any two-particle interaction; hence, the wide variety of physical qubits. 

%\paragraph{special gate sets: Clifford and FF} 
There are two gate sets of special interest because, not only do they fail to span the full set of transformations, they are also tractable with conventional computing. The first is the Clifford gate set\cite{Gottesman1998Jul}. The H, S, and CNOT gates are the generators of the Clifford group. A circuit composed of only the Clifford gates can be computed using conventional computation. Note that the T gate listed in the Table of gates allows for the Clifford set to realize universal quantum and some quantum algorithmic costs estimates are given by the T gate count of the implementation. The second specialized gate set is the matchgate~\cite{valiant2002quantum} (or free-fermion \cite{Terhal2002Mar} gate set) which can also be efficiently simulated without quantum resources. First, we begin with the Clifford gate. 

%\paragraph{special gate sets: Clifford}
In the case of the Clifford gates (H, S, CNOT given in Table \ref{tbl:HPTCNOTgates}), a family of operators (the Pauli group) $\mathcal{P}=\{P_1,P_2,P_3,P_4,P_5,...\}$ is preserved such that action of Clifford gates will only transform each element $P_j$ into another element $P_k$ but not into linear combinations of elements as in $0.4P_{k} + 0.6P_{k'}$. This gate set was studied in the theory of quantum error correction \cite{Gottesman1998Jul, gottesman2010introduction, Girvin2021introduction,nielsen2002quantum} leading to the development of the stabilizer formalism\footnote{The stabilizer formalism extends families of operators, like the Pauli group $\mathcal{P}$, to a corresponding set of vectors $\{\vec\psi_k\}_{k=1}$ that are `stabilized' by that particular family of operators. Here stabilized means that the action of any operator from the family does not change the corresponding set of vectors $\{\vec\psi_j\}$ e.g. in the case only one vector is stabilized: $P_m\vec\psi_0=\vec\psi_0$. If an unwanted error occurs then the resulting vector will leave the stabilized subspace and in many instances, this can be both detected and corrected \cite{Gottesman1998Jul, gottesman2010introduction}.}  widely employed in quantum information science. Note that many of the early ideas for quantum algorithms that relied exclusively on superposition are simulatable with the Clifford formalism e.g. Deutsch's algorithm \cite{deutsch1985quantum}.

%\paragraph{special gates: matchgates and fermion/bosons}
A second set of gates that are also tractable with conventional computing is the matchgate set \cite{valiant2002quantum}. This set is also known as free-fermion gates \cite{Terhal2002Mar}. Both the perfect matching problem\footnote{The perfect matching problem for a graph with given vertices and edges requires selecting a set of edges where every vertex has only one neighbor. The selected edge perfectly matches all vertices into pairs.} for \emph{planar} graphs \cite{barahona1982computational} and the physical description of \emph{non-interacting} fermionic particles%
\footnote{Fermion are identical particles that carry the antisymmetric one-dimensional representation of the symmetric group of permutations. Since this representation is one-dimensional, the action of permutations does not change the quantum probability density matrix. The other one-dimensional representation of the symmetric group irrespective of the number of elements being permuted is the completely symmetric representation which corresponds to bosons discussed further in Section \ref{sec:primacy}.}
are deeply connected to the computational and mathematical properties of determinants.%
%%%%%%
\footnote{For the present discussion, let us recall \cite{Horn2012Oct} that a determinant is characterized by
\begin{equation}
    det(A) = \sum_{\pi\in S_n}^{n!} \text{sgn}(\pi) \prod^n_i A_{i,\pi(i)}
    \label{eq:det}
\end{equation}
where $A$ is an $n\times n$ matrix and $S_n$ is the set of permutations of $n$ objects. %index by $\pi$ and 
Here $\text{sgn}(\pi)$ is $\pm1$ depending if the permutation $\pi$ is composed of an even e.g. $P_{12}P_{23}$ or odd e.g. $P_{12}P_{23}P_{12}=P_{13}$ number of transpositions. This summation over $n$ factorial factors to obtain $det(A)$ is equal to the product of the eigenvalues of $A$. The eigenvalue computation is much faster than summing over the formula in Eq.~\eqref{eq:det}. The same combinatoric function without the $\text{sgn}(\pi)$ is called the permanent: $perm(A)= \sum_{\pi\in S_n}^{n!}  \prod^n_i A_{i,\pi(i)}$. This corresponds to non-interacting particles called bosons \cite{Aaronson2013Feb}.}
%%%%%%
%
This connection enables circuits composed of matchgates to be efficiently computed without the help of quantum resources \cite{Terhal2002Mar,valiant2002quantum}.

%\paragraph{operation speed / coherence time} 
%As we described in the introduction, there are many experimental realities when it comes to constructing a quantum computational.  Unlike conventional computers highly focused on semi-conductor technology, the variety of quantum devices leads to a range of gate time speeds from \changeme{superconductors at XX} and \changeme{other qubit at YY}. In all platforms, the control of the qubits can be characterized by the quality of the qubits; that is, how long they can hold information both quantum and in general. The noise sources which cause decoherence include noise arising from background sources such as air particles bombarding trapped ions containing quantum information and experimental control errors in the microwave pulse sequences used to control superconducting. Similarly, the noise in the two-qubit couplings also plays an important role in understanding device performance and its ability to generate entanglement. 

%\paragraph{Alternatives to gate model [Done]}
So far, we have been discussing quantum operations as a discrete set of gates that are applied in a prescribed sequence. However, the quantum circuit model is not the only way to describe quantum computation just as there are equivalently powerful ways to control a conventional computer e.g.~Fortran or C++.  Similar to the difference between conventional programming languages%
\footnote{One may argue the identity of programming languages mostly lies in their syntaxes and programming paradigms. However, in the current context, two programming paradigms are called equivalent if they both simulate the same class of functions up to a polynomial-time overhead due to the conversion \cite{sipser_introduction_2012}. }%
, some quantum computational models align with certain quantum devices better than others. 

%alternates to the quantum circuit model
Some important alternatives to the quantum circuit model include quantum Turing machines~\cite{Bernstein1997complexity}, adiabatic quantum computation~\cite{farhi2000quantum, aharonov2008adiabatic}, quantum walk models~\cite{childs2009universal}, and dissipative quantum computing~\cite{verstraete2009quantum}.  Additionally, a wide variety of layered circuit designs have been shown to be universal for quantum computation such as quantum approximate optimization~\cite{Farhi2014Nov} and variational quantum eigensolver~\cite{biamonte2021universal}. 
In Section \ref{sec:applications}, we collect these observations using the quantum signal processing framework \cite{gilyen2019quantum}.

\section{Quantum primacy experiments}\label{sec:primacy}

Quantum primacy refers to the situation where a quantum device has performed a computation that cannot be done in any other way. Note that the computational task may or may not have practical significance but is chosen merely to show the separation between quantum and conventional computing devices.

% RCS + BS
The two most widely discussed approaches to quantum primacy~\cite{lund2017quantum,Harrow2017} are the BosonSampling~\cite{Aaronson2013Feb} and Random Circuit Sampling~\cite{boixo2018characterizing} problems. In the first of these problems, Random Circuit Sampling, considers the sampling from the output distribution of randomly generated quantum circuits~\cite{boixo2018characterizing}. It is important to note that these quantum primacy experiments are based on sampling arguments which further underline the tight correspondence between quantum and probability theory discussed earlier. The second problem, BosonSampling, a few photons are sent into a randomly selected linear-optics setup and the computational task is to sample from the output distribution of photons exiting the optical network~\cite{Aaronson2013Feb}. 

% RCS
After the attention received by the first experimental result claiming quantum primacy \cite{arute2019quantum}, there has been a fascinating string of results both experimentally and theoretically leaving the exact boundary of primacy still an open area of active research. The first claim of quantum primacy was accomplished using a 53 transmon qubit device executing and sampling around 1,500 random gates as part of a circuit with 20 layers deep~\cite{arute2019quantum}. However, this was for brute force approaches. Since then, larger experimental implementations of Random Circuit Sampling with more qubits  (56 qubits and with 60 qubits) with a circuit depth of 24 layers~\cite{zhu_quantum_2021, wu_strong_2021}.

%experimental 
Initially, it was claimed \cite{arute2019quantum} that conventional computers would take 10,000 years to accomplish the same task.  This assumed a brute force simulation however this has been vastly improved using tensor networks methods based on the analysis of states with low entanglement \cite{Verstraete2008Mar,Cirac2021rmp}. The application of heavy computational resources and clever entanglement truncation schemes have has led to much improved conventional computational results largely closing the computational gap \cite{gray2021hyper,zhou2002limits,pan_solving_2021,liu2021redefining,kalachev2021recursive}. Regardless of the classical computational progress, the cost of performing tensor network simulations scales exponentially with the number of qubits so these approaches can be defeated by going to larger experimental qubit devices. Note that even if the original quantum primacy circuits are simulated without quantum resources, performing increasingly larger Random Circuit Sampling instances should increase the difficulty beyond conventional computing since tensor networks scaling exponentially with the system size. Theoretically, this idea is supported \cite{lund2017quantum,Harrow2017,aaronson2019classical} based on complexity conjectures. Unfortunately, the standard statistical benchmark used to certify correct RCS, the linear cross entropy, has major deficiencies \cite{gao2021limitations}.

%XEB
The linear cross entropy benchmark was chosen as an easy to estimate proxy for the fidelity of the quantum state prepared. The fidelity, the total variation distance and the cross entropy require exponential samples to estimate while the linear cross entropy does not. However, there are statistical vulnerabilities when the linear cross entropy is used instead of other metrics that require exponential more sampling. More startling is the asymptotic scaling of the linear cross entropy as compared with standard measures like the fidelity. For conventional algorithms attempting to hack or otherwise spoof the linear cross entropy, it can be shown that their performance actually improves with system size. This behavior is drastically different from the scaling of the fidelity between $N$ approximate states and $N$ ideal states asymptotically. \footnote{If an approximate state has overlap $s$ with the ideal state then, the combined state of $N$ non-interacting copies have overlap of $s^N$ which is asymptotically vanishing.} Thus, there is need for a new routes to validating the results of Random Circuit Sampling. 

%BosonSampling
The second route to quantum primacy also under active debate uses the BosonSampling problem \cite{Aaronson2013Feb}.  The BosonSampling problem relies on the hardness of approximating the permanent of a matrix and its deep connection to the nature of indistinguishable bosons.  BosonSampling  can be implemented with linear optics but requires less than full-fledged universal photonic quantum computing \cite{knill2001scheme}. The first claims of quantum primacy using a quantum optics setup appeared in 2020~\cite{zhong_quantum_2020} and have been since been improved~\cite{zhong_phase-programmable_2021}. Similar to the discussion ensuing the Random Circuit Sampling primacy experiment, improved conventional computational methods are claimed to simulate the early experimental BosonSampling primacy results~\cite{villalonga_efficient_2021}. However, the mid-2022 announcement from Xanadu Computing provides a new quantum primacy experimental claim~\cite{madsen2022quantum}.  %Thus, here also the claim of quantum primacy remains elusive.

%outlook
The experimental and theoretical progress in understanding the boundaries between quantum and classical computing is an active and growing area of research. This race between quantum and conventional technologies an interesting area to follow in 2022. Next, we turn our attention toward practical applications and quantum algorithms.

\section{Quantum computing algorithms and applications}\label{sec:applications}

%\paragraph{opening}
Applications of quantum computing are as diverse as the fields necessary to create quantum information processing technology. In recent years, the excitement has attracted further talent and investment sparking a rapid growth in the quantum algorithm development \cite{montanaro2016quantum}.  There now exists a zoo of quantum algorithms \cite{jordanZoo} and, in this section, we hope to provide a taxonomy with for understanding and traversing the literature. 
We do so by dividing our quantum algorithm classification into two overlapping groups: the first wave quantum algorithms based on formal methods and the second wave quantum algorithms largely based on optimization methods. We will connect these two groupings using the quantum singular value transformation framework which broadly and elegantly combines decades of theoretical quantum algorithms research.

%division of ideas
Initial ideas about quantum computation were generated with heavy input from the mathematics and computer science community, where the questions of noise were considered only formally and shown to be theoretically surmountable. Therefore, the first wave of quantum algorithms assumed noiseless quantum device operation (or otherwise fully quantum error corrected systems). The availability of noisy intermediate-scale quantum (NISQ) \cite{preskill2018quantum} computers spurred the second wave of quantum algorithms and quantum tasks that more easily take into account noise and advances in algorithm design on conventional computers. We begin with the first wave, followed by the second wave.

\subsection{First wave quantum algorithms for ideal quantum devices}

%\paragraph{Key quantum algorithm from a CS point of view}
The initial burst of quantum computing algorithms was developed with an ideal quantum computer in mind. 
Quantum Turing models, quantum gate model, and query-based oracle models were studied largely without noise.  
The development of quantum error correction~\cite{gottesman2010introduction, Girvin2021introduction} was pursued alongside the first wave quantum algorithms with sufficient theoretical success that algorithm designers felt they could largely ignore noise and details of the physical architecture. 
The canonical reference for this wave of quantum algorithm development is \cite{nielsen2002quantum} which despite being over 20 years old still stands as a solid reference for the theoretical background of quantum computing and quantum information. The recent review article Ref.~\cite{preskill2021quantum}, also provides a canonical introduction to quantum computing with an up to date perspective on the field.

%summary of section
The two foundational results culminating from these effort are the Grover search and the quantum Fourier transform subroutines. We begin with Grover search.  

%\subsubsection{Grover quantum search algorithm}
Grover search is a quantum algorithm for finding a marked item in an unsorted database of size $N$. This task ordinarily takes a number of trials that is proportional to the size of the database.  There is no particular shortcut other than testing each item in the database for a hit. An apt example is searching for a specific tool in an unsorted shed. Each item can be evaluated as correct or incorrect by the mechanic e.g. who sits far away from the shed. With Grover's quantum algorithm, the marked item can be found in a time that is proportional to the square root of the database size. % $\sqrt{N}$.

%Grover optimality
Grover's search was announced in 1996 \cite{Grover1996Jul} and it was one of the few quantum algorithms that had proof of optimality for the task it solved~\cite{bennett1997strengths}. This means for problems belonging to the complexity class NP, quantum algorithms can, at most, achieve quadratic speedup~\cite{bennett1997strengths}. Informally, this means if the best known computational method for solving a sufficiently generic problem is just the brute-force testing of every possible solution, then optimal use of quantum resources will achieve only a quadratic improvement over the conventional brute-force algorithm. 

%main ideas for grover
To understand the major themes of quantum algorithm development, it is useful to delve deeper into the theory behind the Grover search as it serves as the backbone of most other quadratic speed ups in quantum algorithm development.  The key idea of the Grover algorithm is the use of a product of reflections where\footnote{A reflection matrix is an unitary (or orthogonal) matrix with determinant equal to -1.} $R^2 = 1$. Reflections are built as $R=(2\Pi-1)$ whenever $\Pi$ is a projector $\Pi=\Pi^2$. 

\begin{figure*}
    \centering
    \[
    \begin{quantikz}[thin lines]
%%%%%%%%%%%%%%%%%%%%%%%%
%\qw & \ket{0}   & \qw & \gate{H} & \qw 
%
& \gate[wires=3, nwires={2}]{R_x} \gategroup[3,steps=1,style={dashed,
rounded corners,fill=yellow!20, inner xsep=2pt},
background,label style={label position=below,anchor=
north,yshift=-0.2cm}]{{\sc }}
%\qw
& \gate{H} \gategroup[3,steps=3,style={dashed,
rounded corners,fill=blue!20, inner xsep=2pt},
background,label style={anchor=
north,yshift=0.2cm}]{\sc }
& \gate[wires=3, nwires={2}]{2\hat\rho_0 - I}
& \gate{H}
&\qw
& \gate[wires=3, nwires={2}]{R_x} \gategroup[3,steps=1,style={dashed,
rounded corners,fill=yellow!20, inner xsep=2pt},
background,label style={label position=below,anchor=
north,yshift=-0.2cm}]{{\sc }}
& \gate{H} \gategroup[3,steps=3,style={dashed,
rounded corners,fill=blue!20, inner xsep=2pt},
background,label style={anchor=
north,yshift=0.2cm}]{\sc } 
& \gate[wires=3, nwires={2}]{2\hat\rho_0 - I} 
& \gate{H}
&\qw
&
%\cdots
%& 
%
%&\qw&\meter{}&\cw	  
\\
%%%%%%%%%%%%%%%%%%%%%%%%
%\qw & \cdots   &     & \cdots   &    
&        & \vdots   &      & \vdots&
&        & \vdots   &      & \vdots&
&\cdots
%&&\cdots&
\\
%%%%%%%%%%%%%%%%%%%%%%%%
%\qw & \ket{0}  & \qw & \gate{H} & \qw 
& %&\qw 
& \gate{H} &  & \gate{H}
&\qw
&
& \gate{H} &  & \gate{H}
&\qw
&%\cdots& 
%
%&\qw&\meter{}&\cw
    \end{quantikz}
\]
(a)
\[
\begin{quantikz}[thin lines]
%%%%%%%%%%%
%Yellow box (1)
	& \gate[wires=3, nwires={2}]{e^{i\phi_{j}(2\Pi-I)}}             
	    \gategroup[3,steps=1, style={dashed,rounded corners, fill=yellow!20, inner xsep=2pt}, background,label style={label position=below,anchor=north,yshift=-0.2cm}]{}	
%Blue Box (3)
	& \gate[wires=3, nwires={2}]{U_x} 
	    \gategroup[3,steps=3,style={dashed,rounded corners,fill=blue!20, inner xsep=2pt},background,label style={label position=below,anchor= north,yshift=-0.2cm}]{}
	& \gate[wires=3, nwires={2}]{e^{i\phi_{j+1}(2\widetilde{\Pi}-I)}}
	& \gate[wires=3, nwires={2}]{U_x^\dagger} 
%%%%
&\qw
%Yellow box (1)
	& \gate[wires=3, nwires={2}]{e^{i\phi_{j+2}(2\Pi-I)}}                 
	        \gategroup[3,steps=1, style={dashed, rounded corners, fill=yellow!20, inner xsep=2pt}, background,label style={label position=below,anchor= north,yshift=-0.2cm}]{}	
%Blue Box (3)
	& \gate[wires=3, nwires={2}]{U_x} 
			\gategroup[3,steps=3,style={dashed, rounded corners,fill=blue!20, inner xsep=2pt},background, label style={label position=below,anchor=north,yshift=-0.2cm}]{}
    & \gate[wires=3, nwires={2}]{e^{i\phi_{j+3}(2\widetilde{\Pi}-I)}} 
	& \gate[wires=3, nwires={2}]{U_x^\dag}	
%%%% (2)
			& \qw 
			& \qw 
			%& \multigate{2}{e^{i\phi_2(2\Pi-I)}}	
			%& \qw 
			\\
%%%%%%%%%%%%%%%%%%%%%%%%%%%
			\cdots
		%	& & &
			&
			& & &	
			&
			&
			& & &
			& 
			& \cdots\\
%%%%%%%%%%%%%%%%%%%%%%%%%%%
			\qw
		%	& & &
			&
			& & &	
			&\qw
			&
			& & &
			& \qw 
			& \qw
\end{quantikz}
\]
(b)
    \caption{The Grover search and the quantum singular value transform share conceptual similarities. The Grover search circuit consisting of two reflections shown in Subfig.~(a). The first reflection $R_x$ is constructed using the oracle deciding if a guess is correct or incorrect. The second reflection $2\hat\rho_0-I$ is about the quantum state $\hat\rho_0$ where all qubits are in the $+Z$ state of the Bloch sphere. The pre- and post-fixed Hadamard gates transform the each qubit of the state into the $+X$ state. This yields the uniform superposition.
    In Subfig.~(b), the generalized circuit for the quantum singular value transformation still consists of a product of two reflections. However, these reflection operations are performed with step-dependent control parameters $\phi_j$ about the subspaces identified by either projector $\Pi$ or $\tilde\Pi$. (Note, since $R^2=(2\Pi-I)^2=I$, we have $\exp(i\phi R)=\cos(\phi)I+i\sin(\phi)R$.) The analysis of the possible functions of $U_x$ reachable after $d$ iterations results in the quantum singular value transformation framework \cite{gilyen2019quantum}.
    }
     \label{fig:grover}
 \end{figure*}
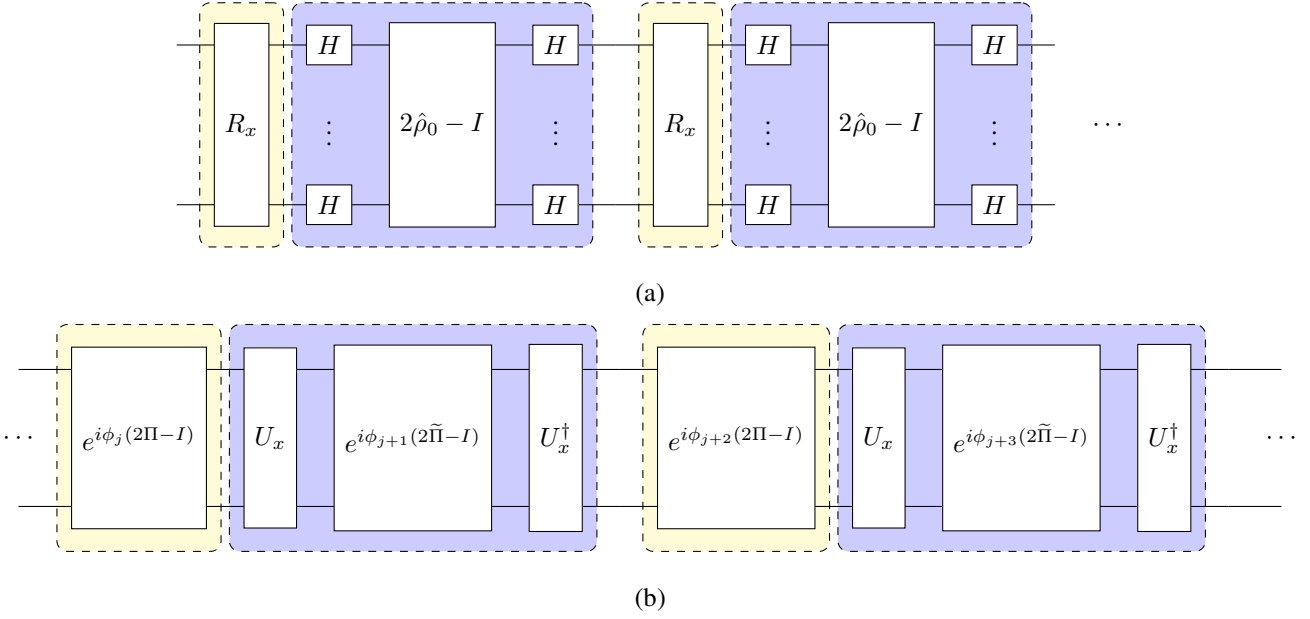

%Product of reflections
The Grover operator, $W=R_AR_x$ is a product of two reflection operators that after approximately $\sqrt{N}$ iterations, transform the initial state $\hat\rho_0$ into a marked state $\hat\rho_x$ with high probability. The first of the two reflection operators, $R_A=2\hat\rho_s-I$, performs a reflection about the pure state generated by applying the Hadamard transformation on all qubits initialized in the Outcome 0 state of the Bloch sphere. The resulting quantum probability density matrix $\hat\rho_s$ is such that all matrix elements including populations and coherences are equal to $1/N$. Second, another reflection operator, $R_x= 2\hat\rho_x -I$, provides phase oracle access to the target pure state, $\hat \rho_x^2=\hat \rho_x$. See Fig.~\ref{fig:grover}a. We will return to this analysis in Section \ref{sec:QSVt} but for now we continue with the quantum Fourier transform and its applications.

%\subsubsection{Quantum Fourier transform}
Quantum factoring and quantum simulation are the two most important applications of quantum computing. Both rely heavily on the quantum Fourier transform.
The quantum Fourier transform (QFT) algorithmic sub-routine improves exponentially over the conventional computing Fourier transform implementations. The exponential separation between these two techniques is at the heart of many of the most exciting applications of quantum algorithms like factoring~\cite{Shor1994Nov}, phase estimation~\cite{Kitaev1995Nov} and the matrix inversion algorithm~\cite{Harrow2009Oct}. 

%factoring
The difficulty of factoring large prime numbers provides modern communication security~\cite{rivest1978method}. The commercial and cryptographic importance of Shor's quantum factoring algorithm sparked much of the excitement and attention of quantum computing in the first wave of quantum algorithm development.  For factoring composite numbers, the quantum algorithm~\cite{Shor1994Nov} requires solving for the periodicity of $f(x)=a^x \mod N$ with $N$ the composite number to be factored, $a$ is a randomly selected integer, and $x \mod y$ equal to the remainder resulting from division of $x$ by $y$.

%time evolution
A similar idea extends the applications of quantum Fourier transform to finding the periodicity of a quantum time-evolution. This is leveraged in quantum simulation tasks ~\cite{Feynman1982Jun, lloyd1996universalqsim} resulting in the phase estimation technique \cite{Kitaev1995Nov}. The phase estimation algorithm allows the eigenvalues of a Hamiltonian to be probed and has found many applications in quantum chemistry and the simulation of physics \cite{Whitfield2010simulation,Georgescu2014qsim, bauer2020quantum, Cao2019}.

\subsection{Second wave algorithms for noisy intermediate-scale quantum devices}

%\paragraph{opening}

%\paragraph{optimization with quantum resources}
This second wave of quantum algorithms attempts to circumvent the use of error correction by relaxing the strict theoretical impositions of quantum computer science \cite{Bernstein1997complexity} and turns towards heuristic optimization procedures. Many of these approaches are designed to take into account the hardware strengths and limitations as well as leverage advances in conventional optimization techniques.

% ansatz methods
The results of quantum computational complexity theory usually focus on worst case analysis \cite{Bernstein1997complexity,kitaev2002classical,Gharibian2015} leaving open the possibility that quantum resources may make progress using heuristic optimization methods \cite{Tilly2021variational,cerezo2021variational}. The resulting wave of approaches typically use feedback between a quantum device and a conventional computer. Error mitigation strategies can be included as part of the quantum optimization loop. In the literature, this family of methods include variational quantum algorithms, variational quantum eigensolvers (especially in the context of quantum simulation problems), quantum approximate optimization algorithm, hybrid quantum-classical algorithms\footnote{These methods highly overlap and often have the same computational power and expressivity. The term \emph{hybrid quantum-classical algorithms} is not a useful way to describe the second wave algorithms because all quantum devices will simultaneously utilize conventional computing to control the quantum device or to dynamically correct errors.} and other anasatz methods. Most of these approaches utilize a conventional computer to perform an optimization procedure using information extracted from the quantum device, usually in an iterative fashion. These quantum optimization methods have been applied to diverse areas such as portfolio optimization \cite{Orus2019} and quantum machine learning \cite{Biamonte2017Sep}. Many of the flagship experiments in quantum optimization were used to find the lowest energy state of the quantum Hamiltonians found in quantum magnets, molecular compounds and other physical simulations contexts, quantum magnets, and other quantum simulation processes~\cite{Whitfield2010simulation,peruzzo2014variational, Kandala2017Sep, OMalley2016Jul,Cao2019,Georgescu2014qsim}. 

% Hamiltonians
Wide interest in quantum computation that is intrinsically robust against errors began with the introduction of the adiabatic quantum computing model~\cite{farhi2000quantum}. In this model quantum instead of thermal fluctuations are used to perform optimization. The first quantum computing company founded, D-Wave Systems, began with this vision in mind. % \changeme{[Gossett]}.
However, the design of the D-Wave devices only allows optimization of Hamiltonian problems of the form
\begin{equation}
H(\vec s)=\sum_{edges: \{ij\}}J_{ij} s_is_j +\sum_{vertices} h_i s_i
\label{eq:H}
\end{equation}
Here $H$, with instances defined by $\{J_{ij},h_i\}$, is a function  of the binary variables $s_i= \pm 1$ e.g. $\vec s =(+1,-1,+1 ...)$. The goal is to find the configuration $\vec s_*$ that minimizes $H(\vec s)$. By varying the parameterization of $H$, we can encode any other constraint satisfaction problems from the computational complexity class NP~\cite{barahona1982computational}. By restricting $s_i$ to be $\pm 1$, the D-Wave devices are only able to solve non-quantum problems with utilizing quantum resources~\cite{ayanzadeh2021multi}. 

% Hamiltonian complexity
One important caveat worth highlighting is that optimization problems can be arbitrarily hard. Thus, many of the claims about quantum optimization must be made heuristically. Because other optimization problems can be mapped to Eq.~\eqref{eq:H}, solving this problem for all instances of $\{J_{ij},h_i\}$ is not likely to be possible with a conventional computer \cite{sipser_introduction_2012}. As noted above one can, at best, achieve a quadratic speed up on this problem. Moreover, when the Hamiltonian in Eq.~\eqref{eq:H} is generalized to a quantum Hamiltonian where the inputs and the constraints can vary with their quantum degrees of freedom e.g.~coherences, the problem only becomes more difficult. The full analysis of worst-case optimizations problems of lowest energy states of quantum Hamiltonians resulted in the theory of Quantum Merlin-Arthur problems~\cite{kitaev2002classical,Gharibian2015}. The takeaway from the results of computer science is that:  even with the promise that quantum resources can verify correct solutions efficiently, generating a targeted specific quantum state is not efficiently doable with or without quantum resources. In other words, just as quantum resources can help solve problems, they can also lead to increased difficulty.

% Scaffolding
So far, we have classified quantum computing algorithms into two branches: those designed for idealized abstractions of quantum computers and the heuristic methods inspired by the quantum hardware currently available. These are not distinct classifications but rather conceptual guidance for navigating the growing literature of quantum algorithms and their applications.  Next, we examine the framework of the quantum singular value transformation. The purpose of including this specific theoretical framework is: (1) it captures the results and ideas from both types of quantum algorithms and (2) the  technical underpinning allows it to subsume many known theoretical results bounding the performance of various quantum algorithms.

\subsection{The quantum singular value transformation framework}\label{sec:QSVt} 
The framework of quantum singular value transformations resulted from vast generalization of known techniques for analyzing quantum algorithms.  The key idea is that a generalized Grover search protocol allows for nearly arbitrary polynomial transformation of a given matrix $A_x$ embedded into the upper left block of a unitary matrix.  Polynomial approximations to $\exp(A_x)$ and $A_x^{-1}$ allows this generalization to subsume not only Grover search but also quantum Fourier transforms, matrix inversion quantum algorithms, quantum adiabatic optimization and many other.  Here, we try to give a clear overview of the structure of the quantum singular value transformation and we refer the reader to~\cite{martyn2021grand, gilyen2019quantum} and references therein for more precise notions. 

%Grover insight
The core insight begins with Grover search. There the quadratic speed up over conventional algorithms is achieved using a product of two reflections as mentioned above. The product of two reflection operators in the case of Grover search results in a two-dimensional subspace where the analysis of the quantum algorithm proceeds. This two-dimensional subspace exists no matter how large the database is and allows one to consider the algorithm using an effective two-dimensional space i.e.~a qubit. This is the essence of the qubitization idea \cite{low2019hamiltonian,szegedy2004quantum}.

% generalized reflections
The first step in the generalization of Grover's search is to allow reflections about more general subspaces than those generated by one dimensional projectors e.g.~pure quantum probability density matrices. This insight was first used to effect in the quantization of pairs of Markov chains~\cite{szegedy2004quantum} where the product of reflections was decomposed into one- and two-dimensional subspaces.

% QSP
The second important realization stems from the mathematical characterization of all possible functions achievable using NMR pulse sequences on a single spin. In NMR, the effective time evolution can be approximated as an alternating sequence of fixed free evolution followed by an experimentally controllable pulse.  For the one qubit situation, this can be described as a fixed rotation about the $Z$-axis of the Bloch sphere, $U_Z(x)$, followed by a parameterized rotation about the $X$-axis $U_X(\phi_j)$. Mathematically, this gives:
\begin{equation}
    W(\vec\phi) = U_Z(x) U_X(\phi_1)
    U_Z(x) U_X(\phi_2)
    %U_z(\theta) U_x(\phi_3)
    ...U_Z(x) U_X(\phi_d)
\end{equation}
The full analytic description of the qubit gates accessible by control parameters $\vec\phi$ to after $d$ alternations is the second major ingredient of the quantum singular value transformation. % Note the similarity 

% polynomial compilation
As may be expected, the resulting final gate is describable using polynomials of degree $d$ of the variable $x$.  
\begin{equation}
    W(\vec\phi)=\begin{pmatrix} 
    poly^{(00)}(x) & poly^{(01)}(x)\\ poly^{(10)}(x) &poly^{(11)}(x) \end{pmatrix}
    \label{eq:w}
\end{equation}
Here each entry is a polynomial with known properties proven in a series of papers all summarised in reference~\cite{gilyen2019quantum}. There now exists numerical algorithms for selecting the parameters to reach target functions \cite{martyn2021grand}.

% Block encoding + QSP + Grover = qSVT
The culmination of these two lines of work resulted in the full analysis of the circuit shown in Fig.~\ref{fig:grover}b. The theorems proven in the quantum singular value framework~\cite{gilyen2019quantum} allows one to extend the understanding of $W(\vec\phi)$  to the transformation of subblocks of unitary matrices. Creating a unitary, $U_x$, with a specific matrix $A_x$ present in the 00 sub-block is called a block encoding of matrix $A_x$. 
$$
U_x=\begin{pmatrix} 
    A_x & \bullet\\
    \bullet &\bullet
\end{pmatrix}
$$
The analysis needed to prove properties of the polynomials in Eq.~\eqref{eq:w} can be extended to prove that the output of the circuit in Fig.~\ref{fig:grover}b will give
\begin{equation}
    W(\vec\phi)=
    \begin{pmatrix} 
    poly^{(00)}(A_x) &\bullet\\
    \bullet &\bullet
    \end{pmatrix}
    \label{eq:w_block_encoding}
\end{equation}

%block encoding and qSVT
The quantum singular value transform 
In Eq.~\eqref{eq:w_block_encoding}, the action of the polynomial on matrix $A_x$ is defined by its action on the singular values of $A_x$ giving the framework its name. The singular value decomposition of $A_x$ is given as
\[
A_x=\sum_k \sigma_k  \; \vec \psi^{L}_k \; \vec\psi_k^{R}
\]
with $\vec \psi^{L}_k$ and $\vec\psi_k^{R}$ the left and right singular vectors corresponding to singular value $\sigma_k$. Then, a polynomial of matrix $A_x$ is given as 
\[
poly(A_x) = \sum_k poly(\sigma_k)\; \vec\psi^{L}_k\; \vec \psi_k^{R}.
\]
The realization of polynomial transformations of the singular values of any block encoded matrix allows the framework to subsume most major quantum algorithms including their asymptotic cost analysis \cite{martyn2021grand,gilyen2019quantum}. 

% Notes
Despite the proofs and analysis of the quantum singular value framework being mathematically involved, we have attempted to provide a useful overview. It should be noted that this section roughly follows the timeline of quantum algorithm development. We started with the first wave that began in the mid-1990's, followed by the crescendo of the second wave activity invigorated by the recent availability of actual quantum devices, then we ended with a powerful framework that will likely seem more activity in the coming years.  

\section{Outlook}\label{sec:outlook}
%Summary of paper
We began with technology, continued with quantum theory and its models of computation.  Now with the overview of quantum algorithms and applications completed, we can add to a few final topics to complete our roadmap to the literature and status update on the field. While maps are useful, deeper appreciation occurs by experiencing the terrain. Therefore, we have ended with a few pointers to resources for accessing quantum devices.

%\paragraph{Accessing quantum computers}
Through the use of cloud computing, companies have already begun offering public access to quantum devices for academic or industrial research purposes. There are mainly two types of services. The first type is cloud services providing access to a single company's collection of quantum devices. The most widely known is the Qiskit cloud services offered by IBM Quantum~\cite{Qiskit}. The second type are multi-platform services that work as intermediaries to give users options to access quantum devices owned by multiple vendors. A key example of such a service is Amazon Braket~\cite{amazonBraketSite} offered through Amazon Web Services. 

%cloud computing
In most cases, the cloud computing interfaces to quantum devices are implemented in Python to provide starting points for accessing working quantum devices. To give an idea of how quantum devices are manipulated, the circuit and corresponding Python snippets for the previously introduced Bell state  (Fig.~\ref{fig:bell}) given in Fig.~\ref{fig:bell-state-circuit}.

\begin{figure}[ht]
\centering
        \begin{quantikz}
 & \gate{H}& \ctrl{1} & \qw \\
 & \qw & \targ{} & \qw
\end{quantikz} 
\begin{minted}[fontsize=\footnotesize]{python}

from qiskit import QuantumCircuit, IBMQ

#Create the Bell-state circuit 
# with 2 qubits and 2 bits
qc = QuantumCircuit(2,2)
qc.h(0)
qc.cx(0,1) 
qc.measure_all()

IBMQ.load_account()
provider = IBMQ.get_provider(
                     group='deployed')
backend = provider.backend.ibmq_manila
job = execute(qc, backend, shots=1000)
########################################
from braket.circuits import Circuit
from braket.aws import AwsDevice

qc = Circuit().h(0).cnot(0,1)

#select a device or simulator
#Rigetti qubit device 
device = AwsDevice(
 "arn:aws:braket::device/qpu/rigetti/Aspen-9")
# IonQ quantum device
device = AwsDevice(
 "arn:aws:braket::device/qpu/ionq/ionQdevice")

result = device.run(qc,shots=1000).result()
\end{minted}
\caption{The quantum circuit and sample Python code for IBM Qiskit and the Amazon Braket service to create the Bell state described in Fig~\ref{fig:bell}. Note that although these commands are valid, these code snippets will not run without properly initializing an account with the correct permissions and running the proper commands to load those credentials.}
    \label{fig:bell-state-circuit}
\end{figure}

%\paragraph{resources}
Finally, we would like to highlight a few resources for learning the implementation of quantum algorithms. 
%Qiskit
The IBM Qiskit textbook~\cite{qiskit_textbook} provides an college-level introduction to quantum information with integrated programming exercises within the text.
%ML + QM see [Pennylane] tutorials~\cite{pennyLane}.
Similarly, the Codebook~\cite{xanadu21} by Xanadu 
provides an introductory course built 
around the Pennylane package providing
differentiable programming of quantum computers. 
%[QBraid.com] tutorials
QBraid is an online platform for developing quantum software that sponsors quantum hackathons and offers introductory quantum tutorials~\cite{qBraid}.  There are many other introductory resources and coding platforms available through public, academic, and commercial channels.

%\paragraph{Closing prayer} 
Armed with the knowledge and perspectives of this review, we hope that the reader is prepared to appreciate and follow this growing area of interest. We have tried to provide a useful starting point for interested graduate students in science and engineering. There are many exciting things to look forward to in the area of quantum computing in 2022. This includes new quantum devices coming online, demonstration of definitive quantum primacy and of practical quantum advantage, stable quantum memory maintained by dynamical quantum error correction and the integration of multiple types of qubits into a single platform.  We are looking forward to both watching and to participating in the evolution of quantum technology and this paper is our invitation to the reader.

\section{Acknowledgements}
We would like to thank  K. Setia, K. Stewart, T. Takeshita, and S. Gulania for useful discussions.  Thanks to U. Vazirani for point out key quantum primacy references. This work was supported by the US NSF (PHYS-1820747, EPSCoR-1921199) and by the Office of Science, Office of Advanced Scientific Computing Research under programs Fundamental Algorithmic Research for Quantum Computing and Optimization, Verification, and Engineered Reliability of Quantum Computers project. This paper was partially supported  by the ``Quantum Chemistry for Quantum Computers'' project sponsored by the DOE. JDW holds concurrent appointments at Dartmouth College and as an Amazon Visiting Academic. This paper describes work performed at Dartmouth College and is not associated with Amazon.

This paper is released under under a Creative Commons Attribution 4.0 International License \cite{CC}.

% \subsection{Quantum devices usage beyond computing}
% \begin{itemize}
% \item Random number generation [Aaronson] (\readmore)
% \item Bad qubits = good sensors: Magnetometers [NV center review]
% \item Time keeping and atomic clocks. [NIST review]
% \item Quantum communication, cryptography. [Companies or at least a review]
% \item Long distance entanglement networks (``flying qubits'')
% \item MSFT quantum inspired imaging [\readmore, press release]
% \end{itemize}

\bibliography{main}{}

\end{document}